\documentclass{aa}
\usepackage{graphicx}
\def\der{{\rm d}}
\begin{document}
\thesaurus{02         
              (12.03.2;  
               12.03.4)}  
   \title{Classification of multifluid CP world models}
   \author {Jens Thomas \and Hartmut Schulz}
\institute{Astronomisches Institut der Ruhr-Universit\"at, D-44780 Bochum, Germany\\
jthomas, hschulz@astro.ruhr-uni-bochum.de \\} 
\date{Received:FILL IN; accepted:FILL IN}
   \maketitle
\markboth{J. Thomas \& H. Schulz: Multifluid world models}
{J. Thomas \& H. Schulz: Multifluid world models}
\begin{abstract}
Various classification schemes exist for homogeneous and isotropic (CP) world models, 
which include pressureless matter (so-called {\em dust}) and Einstein's cosmological
constant $\Lambda$. We here classify the solutions of more general world models
consisting of up to four non-interacting fluids, each with pressure ${\cal P}_j$,
energy density $\epsilon_j$ and an equation of state 
${\cal P}_j = (\gamma_j - 1) \epsilon_j$ with $0 \le \gamma_j \le 2$. 

In addition to repulsive fluids with negative pressure and positive energy density, which
generalize the classical repulsive (positive) $\Lambda$ component, we 
consider fluids with negative energy density as well. The latter generalize a negative
$\Lambda$ component. This renders possible new types of models that do 
not occur among the classical classifications of world models. Singularity-free periodic solutions as well as
further `hill-type', `hollow-type' and `shifting-type' models are feasible. 

However, if one only allows for three components (dust, radiation and one repulsive component)
in a spatially flat universe
the repulsive classical $\Lambda$ fluid (with $\Lambda > 0$) tends to yield the smoothest 
fits of the Supernova Ia data from 
Perlmutter et al. (1999). Adopting the SN Ia constraints, exotic negative 
energy density components can be fittingly included only if the universe consists of four or more fluids.
\end{abstract} 
%
\section{Introduction}
While in the Newtonian description gravity is only considered as an
attractive force, repulsive effects are possible in general relativity, as
e.g.\ caused by Einstein's (1917) cosmological constant $\Lambda$ (if positive). 
The variety of general-relativistic
CP world models (CP=cosmological principle which requires homogeneity
and isotropy after averaging on suitable scales; we require exact homogeneity and isotropy
for the mathematical CP models) including the $\Lambda$-term and 
non-relativistic incoherent, comoving (i.e.\ with pressure 
${\cal P}=0$) matter (so-called {\em dust}) was classified using different
parametrizations by Robertson (1933), Stabell \& Refsdal (1966) and Carroll et al. (1992).
From the seventies till the nineties, a widely followed assumption was to set  $\Lambda=0$, partly promoted by the
inability of current quantum field theory to predict a realistic value of $\Lambda$ 
(e.g. Weinberg 1989). 

This attitude changed during the last few years. The near-flatness suggested by cosmic
microwave background (CMBR) observations (Lineweaver 1998, de Bernardis et al. 2000) 
and the time-scale problem caused by a tendency to a large Hubble constant 
led an increasing number of workers to consider world models with $\Lambda$ or similarly repulsive fluids 
(e.g. Ostriker \& Steinhardt 1996; Turner \& White 1997). Recently, direct motivation for such an approach has come 
from observations of supernovae of type Ia, taken as cosmic
`standard candles', that require the presence of an accelerating component in the expanding 
universe (Riess et al.\ 1998; Perlmutter et al.\ 1999), primarily taken as
evidence for the action of $\Lambda$. If so, CMBR data and
peculiar motions of galaxies can in principle break the degeneracy between the decelerating
and accelerating components given by the SN data alone, so that it may become feasible 
to determine an accurate value of $\Lambda$ (Zehavi \& Dekel 1999).

However, given our ignorance about the physical origin of $\Lambda$ it may be premature to identify 
the $\Lambda$ term with the repulsive component. E.g., in the early universe inflaton fields
or domain walls are well known to serve as expansion-driving forces as well (Kolb \& Turner 1990).
As will be recalled in Sect.\,\ref{basic}, any ideal fluid with negative pressure may force
expansion of space if its equation of state obeys a certain condition. 

It is the goal of the present work to extend the above cited classifications of CP world 
models in two ways. 
Firstly, we allow for {\em more than two} components.
Secondly, we allow for a range of fluids which have in common
that each equation of state yields proportionality between energy density $\epsilon$ and
pressure $\cal P$. This form of the equation of state includes the classical cosmological
matter components dust, radiation and the $\Lambda$ fluid, but also encompasses more general repulsive and
attractive components. In addition to the above mentioned negative-pressure components
(generalizing the $\Lambda > 0$ case)
we also allow for fluid components with negative energy densities which generalize the case
of a negative $\Lambda$. 
The present work extends a discussion started in Schulz (1998).
\section{Basic equations}
\label{basic}
To fix our notation we now compile the basic equations for the subset of CP world models
to be studied here. The vacuum velocity of light is set $c=1$.
Firstly, we recall that the $\Lambda$ term is equivalent to
a fluid with constant energy density $\epsilon_{\Lambda}$.

\subsection{Constant-density fluid}
\label{constdens}
Einstein's field equation in the general form reads
\begin{equation}
\label{field}
R_{\mu\nu} - \frac {1}{2} R\, g_{\mu\nu} + \Lambda g_{\mu\nu}= -8 \pi G\, T_{\mu\nu}
\end{equation}  
in which we take as source term on the right side a sum
over ideal-fluid energy-momentum tensors of the form 
\begin{equation}
\label{Tfluid}
T^{\mu\nu}_{\rm id. fluid} = (\epsilon + {\cal P})\,u^{\mu} u^{\nu} - {\cal P}\,g^{\mu\nu}.
\end{equation}
${\cal P}$ and $\epsilon$ are pressure and energy density measured in a local comoving frame, in
which the four-velocity reduces to $u^0=1$, $u^1=u^2=u^3=0$. Despite lacking a definition
of their four-velocity,
isotropic relativistic components (null-fluids) can be subsumed under this scheme
with an equation of state ${\cal P}=\epsilon/3$. 
The $\Lambda$ term in Eq.(\ref{field}) can be omitted on the left side and instead, on the right side,
be represented by a `fluid' after substituting $\Lambda = 8 \pi G \, \epsilon_{\Lambda}$
with a constant energy density $\epsilon_{\Lambda}$ and adopting an equation of state 
${\cal P}_{\Lambda} = - \epsilon_{\Lambda}$ for the $\Lambda$-fluid.

\subsection{Fluid equations}
The basic symmetries given by the assumption of homogeneity and isotropy
yield the Robertson-Walker (RW) line element 
which reads with spatial polar coordinates $x^0=t, x^1=\sigma, x^2=\theta,
x^3=\phi$
\begin{eqnarray}
\label{RW}
\der \tau^2 & = &g_{\mu\nu} \der x^{\mu} \der x^{\nu} = \nonumber \\
  & = & \der t^2 - a^2(t) \left ( \frac {\der \sigma^2} {1-k\,\sigma^2} +
\sigma^2 (\der \theta^2 + \sin^2 \theta\, \der \phi^2)\right)
\end{eqnarray}
$k \in \{0,-1,1\}$ is a curvature index. As is common practice we call models with $k=0, -1,1$
flat, open or closed, respectively, without emphasizing its relation to spatial curvature.
Here $\sigma$ is fixed for comoving fluid elements. 
The right side of Eq.\,(\ref{field}) is subject to the conservation law
for matter fields $T^{\mu\nu}{}_{;\mu}=0$ 
yielding the equation of continuity for perfect fluids
$$u_{\nu}\,T^{\mu\nu}{}_{;\mu} = (\epsilon u^{\mu})_{;\mu}+{\cal P}\, u^{\mu}{}_{;\mu}=0$$
which reduces with the RW line element to the cosmological energy equation 
\begin{equation}
\label{ener}
{\dot \epsilon}  = - 3 \,(\epsilon + {\cal P}) \, \frac {\dot a} {a} \ \ \
\mbox{or} \ \ \ 
\frac {\der (\epsilon a^3)} {\der a} = - 3\, {\cal P} \, a^2. 
\end{equation}
The relativistic Euler equation (also stemming from $T^{\mu\nu}{}_{;\mu} = 0$)
\begin{equation}
\label{euler}
(\epsilon +{\cal P})\,u^{\mu} \, u^{\nu}{}_{;\mu} =(g^{\mu\nu} - u^{\mu}\,u^{\nu})\, {\cal P}_{;\mu}
\end{equation}
reduces to $u^{\mu} \, u^{\nu}{}_{;\mu} = 0$, showing that an 
ideally homogeneous and isotropic universe has geodesic flow lines.

The relativistic analogue of Newtonian mass density in the
Euler equation Eq.\,(\ref{euler}) is 
$(\epsilon + {\cal P})$ and can therefore be considered as generalized {\em inertia}.
For the stability of `normal' matter $(\epsilon + {\cal P}) > 0$ is required. Fluctuations in the 
spatial distribution of energy cause deviations from geodesic flow, if 
$(\epsilon + {\cal P}) \neq 0$. For the $\Lambda$-fluid, because of its vanishing inertia,
inhomogeneities are inconsistent with Eq.\,(\ref{euler}).

Inserting the RW line element Eq.\,(\ref{RW}) into the field equation Eq.\,(\ref{field})
leads to the Friedmann equations
\begin{equation}
\label{fried1}
\frac {\ddot{a}} {a}  =  - \frac {4 \pi G} {3} (\epsilon + 3 {\cal P})
\end{equation} 
and
\begin{equation}
\label{fried2}
H^2(t) \equiv \left( \frac {\dot{a}} {a} \right)^2 = \frac {8\pi G} {3} \epsilon - \frac {k}{a^2}
\end{equation}
$H(t)$ is the Hubble parameter at time $t$.
The system of Eqs.\,(\ref{ener}), (\ref{fried1}) and (\ref{fried2}) is redundant; from any two of these
equations one can derive the third. 
\subsection{Equation of state}
For full specification of a gravitating source one needs an equation of state.
Here we assume for each cosmic fluid 
\begin{equation}
\label{eqs}
{\cal P} = (\gamma - 1) \epsilon
\end{equation}
where $\gamma$ is a constant specifying the fluid. This form was chosen by
Zeldovitch (1962) as an equation of state also common for nuclear matter.
It applies to the best known cosmic fluids like (e.g. Kolb \& Turner 1990)
\begin{itemize}
\item dust (${\cal P}=0; \gamma=1$),
\item radiation and any other extremely relativistic matter (${\cal P}=\epsilon/3; 
\gamma=4/3$), 
\item the constant-density $\Lambda$-fluid (${\cal P}=-\epsilon; \gamma=0$), 
\item a network of slow cosmic strings (${\cal P}=-\epsilon/3; \gamma=2/3$) or 
\item domain walls (${\cal P}=-(2/3)\epsilon; \gamma=1/3$). 
\end{itemize}
Apart from incoherent matter (dust) and radiation the physical 
background of cosmic fluids with equation of state\,(\ref{eqs}) is hardly understood.
Therefore, in case of $\gamma \ne 1$ or $4/3$ we use the designation {\em exotic fluid}.

We confine the parameter of the equation of state to the range
\begin{equation}
\label{gam}
0  \leq  \gamma \leq 2 \ \ \mbox{implying} \ \ c_s \leq c
\end{equation}
where the speed of sound $c_s$ is restricted not to surpass the 
velocity of light $c$.

Inserting Eq.\,(\ref{eqs}) into Eq.\,(\ref{ener}) and integrating yields
\begin{equation}
\label{eps}
\epsilon = \epsilon_0 \left( \frac {a} {a_0} \right)^{-3\gamma}
\end{equation}
where the constants $\epsilon_0$ and $a_0$ are here fixed at 
$t=t_0$, the present epoch. 

For the $\Lambda$-fluid  $(\epsilon_{\rm \Lambda} + {\cal P}_{\rm \Lambda}) = 0$ Eq.\,(\ref{ener}) yields 
$\dot{\epsilon}_{\rm \Lambda}=0$
or $\epsilon_{\rm \Lambda} = \mbox{const}$ as required in Sect.\,\ref{constdens}. Eq.\,(\ref{euler}) rules out any
pressure gradient in the $\Lambda$-fluid. Hence, the density of a universe consisting solely of the $\Lambda$-fluid
would be {\em spatially and temporarily} constant, i.e. it would satisfy the {\em Perfect Cosmological
Principle} of Bondi \& Gold (1948) and Hoyle (1948). Indeed, McCrea (1951) derived the equation of state
$\cal P = - \epsilon$ for the matter content of the steady-state universe.

Since the definition of the cosmological constant by Einstein (1917) positive as well as
negative values of $\Lambda$ have been considered. Transferring this to the
fluid description given in Sect.\,2.1  suggests to allow for positive {\em and}
negative energy densities of other $\gamma$ fluids in our phenomenological approach as well. 

Static world models like Einstein's (1917) historic solution will not be included 
in the
classification diagrams below (albeit mentioned in Sect.\,\ref{bassol}). The FLRW 
interpretation of the redshifts
of galaxies makes such models obsolete. In our context, de Sitter´s (1917) 
historic world model
corresponds to a single-fluid universe with $\gamma=0$. Despite 
misleading information in some textbooks, Friedmann's (1922, 1924) solutions
are two-component models consisting of dust ($\gamma = 1$) and the $\Lambda$ 
fluid ($\gamma=0$)
(as has been noted by Felten \& Isaacman (1986) as well).

\subsection{The multifluid approach}
\label{multfluid}
We assume a finite number of non-interacting\footnote{Non-gravitational interactions,
i.e. interactions other then through the Friedmann equations, are excluded.} cosmic fluids
with equations of state of the form Eq.\,(\ref{eqs}),
an energy equation of the form (\ref{ener}) and, hence, Eq.\,(\ref{eps}) holds for each fluid
so that
\begin{equation}
\label{mener}
\epsilon = \sum_i \epsilon_i \ \ \ \ {\cal P} = \sum_i (\gamma_i -1) \epsilon_i \ \ \ 
\mbox{with} \ \ \
\epsilon_i = \epsilon_{0i} \left( \frac {a} {a_0} \right)^{-3\gamma_i}
\end{equation} 
Defining the normalized density $\Omega_i$, the normalized scale factor $x(t)$ and the parameter of state
$\alpha_i$ by the equations
$$
\Omega_i := \frac {8\pi G} {3 H_0^2} \epsilon_{0i}, \ \ x(t) := \frac{a(t)}{a_0} \ \ {\rm and} \ \ 
\alpha_i := 2-3\gamma_i
$$
we can write Eqs.\,(\ref{fried1}), (\ref{fried2}) in the form
\begin{equation}
\label{exp1}
\ddot{x} = \frac {H_0^2} {2 x} \sum_i \alpha_i \Omega_i x^{\alpha_i} =: f_1(x)
\end{equation}
\begin{equation}
\label{exp2}
\dot{x}^2 = H_0^2 \left(\sum_i \Omega_i x^{\alpha_i} +1 - \sum_i \Omega_i \right) =: f_2(x).
\end{equation}
In these equations we have introduced the abbreviations $f_1(x)$ and $f_2(x)$ for the right-hand sides
of the Friedmann-equations. Therebye we emphasize that $f_1$ and $f_2$
are explicit functions of the scale factor $x$, which only mediate the dependence on cosmic time $t$.
It is mathematically crucial for the classification that $f_1(x)$ and  $f_2(x)$ do {\em not explicitely} depend on time.
We also want to distinguish between the different role of the dynamical functions $\dot{x}$ and $\ddot{x}$
and the conditions $f_1(x)$ and $f_2(x)$ that determine the temporary behavior of the scale factor $x(t)$.
 
To derive Eq.\,(\ref{exp2}) one has to use the present-epoch ($t=t_0$) version of Eq.\,(\ref{fried2}) to eliminate the
curvature index $k$ of Eq.\,(\ref{RW})
\begin{equation}
\label{curv}
k = a_0^2 H_0^2 \left( \sum_i \Omega_i - 1 \right).
\end{equation}

By differentiating Eq.\,(\ref{exp2}) with respect to $x$ one obtains
\begin{equation}
\label{diffried}
\frac{1}{2} \frac{\der f_2(x)}{\der x}=\frac{1}{2} \frac{\der \dot{x}^2(x)}{\der x}=\ddot{x}(x)=f_1(x), 
\end{equation}
which is an important property to be used below.

Eq.\,(\ref{exp2}) can be formally integrated by
\begin{equation}
\label{world}
H_0 (t-t_{\rm init}) = \int_{x_{\rm init}}^x 
\frac {\der x}{\sqrt{\sum_i \Omega_i x^{\alpha_i}
 + 1 - \sum_i \Omega_i}}.
\end{equation}
In big-bang models the initial time $t_{\rm init}=0$ can be chosen for the `big bang'
(given by the first instant with $x = 0$ when progressing backwards in time). In models 
with an infinite past $t_{\rm init}$ can be chosen in accordance with one's wish how far to follow the model.
Eq.\,(\ref{world}) gives $t(x)$ which by inversion (sometimes of several
monotonic branches) leads to the desired world model $x(t)$. 

In practice, rather than carrying out the direct integration it is frequently
more convenient to solve Eq.\,(\ref{fried1}) via a standard fourth-order Runge-Kutta routine.
Good examples for the computation of the case $\Lambda$ + dust models are given by
Felten \& Isaacman (1986).
\section{Cosmic dynamics}
\label{dynamics}
In the following we call components with $\alpha > 0$ {\em $\Lambda$-like}, those with $\alpha < 0$ 
{\em dustlike} because of their dynamical {\em similarity} to the $\Lambda$-fluid ($\alpha = 2$) and 
cosmic dust ($\alpha = -1$), respectively.
\subsection{Boundaries of the scale factor}
\label{extr}
The physical meaning of $x(t)$ requires
\begin{equation}
\label{conditions}
\dot{x}^2 \geq 0 
\end{equation}
restricting $\dot{x}^2$ to the positive branch of $f_2(x)$,
so that $\dot{x}$ and $x$ are real valued. We assume $x \geq 0$ and identify the big-bang with $x = 0$.
As a consequence of Eqs.\,(\ref{ener}) and (\ref{eqs}), $f_1$ and $f_2$ (cf. Eqs.\,(\ref{exp1}), (\ref{exp2})) 
depend on $x$ and  only indirectly
on cosmic time $t$ via $x(t)$ (as pointed out in Sect.\,2.4). Thus, {\em geometric} features of the 
solution $x(t)$ are naturally related to $x$ rather than to $t$.
For instance, if at a certain time $t_{\rm max}$ the cosmic evolution passes a maximum 
$x_{\rm max} = x(t_{\rm max})$, then for any other maximum at $t'>t_{\rm max}$ necessarily
$x(t') = x_{\rm max}$ holds, meaning that $x_{\rm max}$ is a global and unique maximal value of
$x(t)$. Obviously the same holds for a minimum $x_{\rm min}$. Consequently, there 
is at most one maximal value of $x$
or at best only one minimum, which we denote by $x_{\rm max}$ and $x_{\rm min}$, respectively. 
It is nevertheless possible that $x(t)$ 
obtains these values infinite times (in periodic models).

For the classification of solutions it is appropriate to distinguish between two different types of 
maxima or minima. Firstly, let us consider a maximum of the scale factor which
is a {\em simple} (or first-order) zero $x_{\rm max}>1$ of $f_2(x)$ by requiring (cf. Eq.\,(\ref{diffried}))
\begin{equation}
\label{maxi}
f_2(x_{\rm max}) = 0, \ f_1(x_{\rm max}) < 0.
\end{equation}
We choose an initial instant of time $t_i$ with $x_i = x(t_i)$, at which the universe is 
assumed to expand. During subsequent evolution the universe
reaches the maximum at time 
\begin{equation}
\label{maxtime}
t_{\rm max} = t_i + \int_{x_i}^{x_{\rm max}} \frac {\der x}{\sqrt{f_2(x)}}.
\end{equation}
Extracting the simple zero of $f_2(x)$ we define a function $g(x)$ by
\begin{equation}
f_2(x) = \left( x_{\rm max} - x \right) g(x),
\end{equation}
so that according to Eqs.\,(\ref{conditions}), (\ref{maxi})
$$
g(x) > 0 \ \ {\rm for} \ \ x_i<x<x_{\rm max}.
$$ 
Thus,  
$$
c := \sup_{x \in \left[x_i,x_{\rm max} \right]} \frac {1}{\sqrt{g(x)}}
$$ 
is finite and
\begin{eqnarray}
t_{\rm max}-t_i&=&\int\limits_{x_i}^{x_{\rm max}} \frac {\der x}{\sqrt{x_{\rm max}-x}\sqrt{g(x)}} < c \int\limits_{x_i}^{x_{\rm max}} \frac {\der x}{\sqrt{x_{\rm max}-x}} \nonumber \\
& & = 2c\sqrt{x_{\rm max}-x_i}. \nonumber
\end{eqnarray}
Hence, in this case
$x_{\rm max}$ will be reached after a finite duration of time and 
may therefore be designated as a {\em maximum} (in contrast to the asymptotic
upper boundary described below). After this finite $t_{\rm max}$
the universe has to contract. World models may exist which pass through such a maximum infinite times.

Analogously $x_{\rm min}<1$ with 
$$
f_2(x_{\rm min}) = 0, \ f_1(x_{\rm min}) > 0
$$
will lead to a {\em minimum} in the above sense (that could occur before $t_i$).

A further possibility is a value $x_{\rm up}>1$ 
with the properties
\begin{equation}
\label{aub}
f_2(x_{\rm up}) = 0 \ \ {\rm and} \ \ f_1(x_{\rm up}) =0.
\end{equation}
Again, from an initial scale factor $x_i = x(t_i)$, where the universe is assumed to expand,
it will reach $x_{\rm up}$ at $t_{\rm up}$ in accordance with Eq.\,(\ref{maxtime}). 
However, when following the above steps and defining a function
$g(x)$ such that 
$$
c := \inf_{x \in \left[x_i,x_{\rm up} \right]} \frac{1}{\sqrt{g(x)}}
$$ 
is finite, we now obtain
\begin{equation}
f_2(x) = \left( x_{\rm up} - x \right)^n g(x),
\end{equation}
with $n \geq 2$. Together with Eq.\,(\ref{maxtime}) this yields
\begin{eqnarray}
\label{maxias}
t_{\rm up}-t_i&=&\int\limits_{x_i}^{x_{\rm up}} \frac {\der x}{\left(x_{\rm up}-x \right)^{\frac{n}{2}}\sqrt{g(x)}} \nonumber \\
&>&c\int\limits_{x_i}^{x_{\rm up}} \frac {\der x}{\left(x_{\rm up}-x \right)^{\frac{n}{2}}}>c\int\limits_{x_i}^{x_{\rm up}} \frac {\der x}{x_{\rm up}-x},
\end{eqnarray}
where the last inequality holds at least in an appropriate neighborhood of $x_{\rm up}$. 
The last integral in (\ref{maxias})
diverges and, consequently, $x_{\rm up}$ will be reached only asymptotically after an {\em infinite} 
time, meaning $t_{\rm up} \to \infty$. 

Furthermore, the scale factor approaches a quasistatic limit at $x_{\rm up}$.
To show this we apply Leibniz' rule on
$$
\frac {\der \ddot{x}}{\der t} = \frac {\der \ddot{x}}{\der x} \dot{x}.
$$
With $h^{(n)} := \frac {{\der}^nh}{\der t^n}$ one obtains
$$
\frac {\der^{n+1} \ddot{x}}{\der t^{n+1}} = \left( \frac{\der \ddot{x}}{\der x} \dot{x} \right)^{(n)}=
\sum_{k=0}^{n} \left({n \atop k}\right) \left(\frac{\der \ddot{x}}{\der x}\right)^{(n-k)} (\dot{x})^{(k)}
$$
which can be written as
\begin{equation}
\label{leibniz}
x^{(n+3)}=\sum_{k=0}^{n} \left({n \atop k}\right) \left(\frac{\der \ddot{x}}{\der x}\right)^{(n-k)} (\dot{x})^{(k)}.
\end{equation}
So, with vanishing 
$$
f_2(x_{\rm up}) = 0 \ \  {\rm and} \ \ f_1(x_{\rm up}) = 0
$$ 
all higher time derivatives of
$x(t)$ vanish at $x_{\rm up}$. (Note that $\ddot{x}$ and its derivatives are continuous for $x>0$.)

In the case corresponding to Eq.\,(\ref{aub}) we call 
$x_{\rm up}$ an {\em asymptotic upper boundary}. Analogously if $x_{\rm low} < 1$ and 
$$
f_2(x_{\rm low}) = 0, \ f_1(x_{\rm low}) =0
$$
then we call $x_{\rm low}$ an {\em asymptotic lower boundary}. 
The universe can begin or end with such states.

It is worth mentioning that due to Eq.\,(\ref{diffried})  
a maximum (or a minimum) of
$x(t)$ is obtained where $f_2(x)$ has a {\em simple} zero. 
An asymptotic boundary is obtained by a {\em higher-order} zero. 

Since the distinction is clear from the context we use below the designations $x_{\rm min}$
for $x_{\rm min}$ {\em and} $x_{\rm low}$ and $x_{\rm max}$ for $x_{\rm max}$ {\em and} $x_{\rm up}$.

The interval within which $x(t)$ is bounded for all possible times $t$ is the 
interval $\left[x_{\rm min},x_{\rm max} \right]$ that
comprises unity and is limited by the first zero $0<x_{\rm min}<1$ of $f_2(x)$  
($x_{\rm min} := 0$, if no such
zero exists) and the first zero $x_{\rm max} > 1$, respectively ($x_{\rm max} := \infty$, 
if no such zero exists).
\subsection{Basic solution types}
\label{bassol}
We first assume a currently expanding universe, i.e. $H_0 > 0$ (as the real universe appears
to be). Depending on whether 
$f_2$ has zeros of the above described types or not, we can
distinguish {\em nine basic types of worldmodels}:
\begin{enumerate}
\item{There is no zero of $f_2(x)$ with $0<x_{\rm min}<1$.}

Such a universe starts with a big bang because, when progressing backwards in time,
its slope $\dot x$ stays positive so that eventually $x=0$ will be reached. If there exists
\begin{enumerate}
\item \label{big bang}no zero $x_{\rm max}>1$ of $f_2(x)$, then the universe expands forever and 
the scale factor grows to infinity;
\item \label{deSimax}a higher-order zero $x_{\rm max}>1$ of $f_2(x)$, then 
the expansion will stop after an infinite time
and the universe becomes asymptotically static with $x \to x_{\rm max}$;
\item \label{big bang + max}a simple zero $x_{\rm max}>1$ of $f_2(x)$, then $x(t)$ will recollapse to $x=0$ 
after passing the maximum $x_{\rm max}$.
\end{enumerate}

\item{$f_2(x)$ has a higher-order zero $0<x_{\rm min}<1$.}

In this case the universe starts with an asymptotically static state at $x = x_{\rm min}$. If there exists 
\begin{enumerate}
\item \label{deSi}no zero $x_{\rm max}>1$ of $f_2(x)$, then the universe expands forever and the scale factor grows
to infinity;
\item \label{shift}a higher-order zero $x_{\rm max}>1$ of $f_2(x)$, then the universe becomes asymptotically static 
after infinite time with the asymtotic scale value $x \to x_{\rm max}$;
\item \label{hill}a simple zero $x_{\rm max}>1$ of $f_2(x)$, then $x(t)$ reaches a maximum and after that decreases. The
contraction stops after an infinite time with $x \to x_{\rm min}$ and the universe becomes asymptotically static.
\end{enumerate}
All these models would have begun in the infinite past.
\item{There is a simple zero $0<x_{\rm min}<1$.}

The universe has not started with a big bang. If there exists
\begin{enumerate}
\item \label{bouncing}no zero $x_{\rm max}>1$ of $f_2(x)$, then the universe  contracts from infinity, passes a minimum
and expands back to infinity;
\item \label{hollow}a higher-order zero $x_{\rm max}>1$ of $f_2(x)$, then the universe starts with the asymptotically
static state $x = x_{\rm max}$, contracts to a minimal value $x_{\rm min}$, expands again and after infinite times
reaches the asymptotic value $x_{\rm max}$;
\item \label{oscill}a simple zero $x_{\rm max}>1$ of $f_2(x)$, then $x(t)$ is periodic.
\end{enumerate}
Because there is no big bang beginning, such world models are infinitely old.
\end{enumerate}

Solutions with at most one boundary are well 
known from the classical dust models with Einsteins cosmological constant $\Lambda$ and 
can be found in Stabell \& Refsdal (1966).

\begin{figure}
\resizebox{\hsize}{!}{\includegraphics[angle=-90]{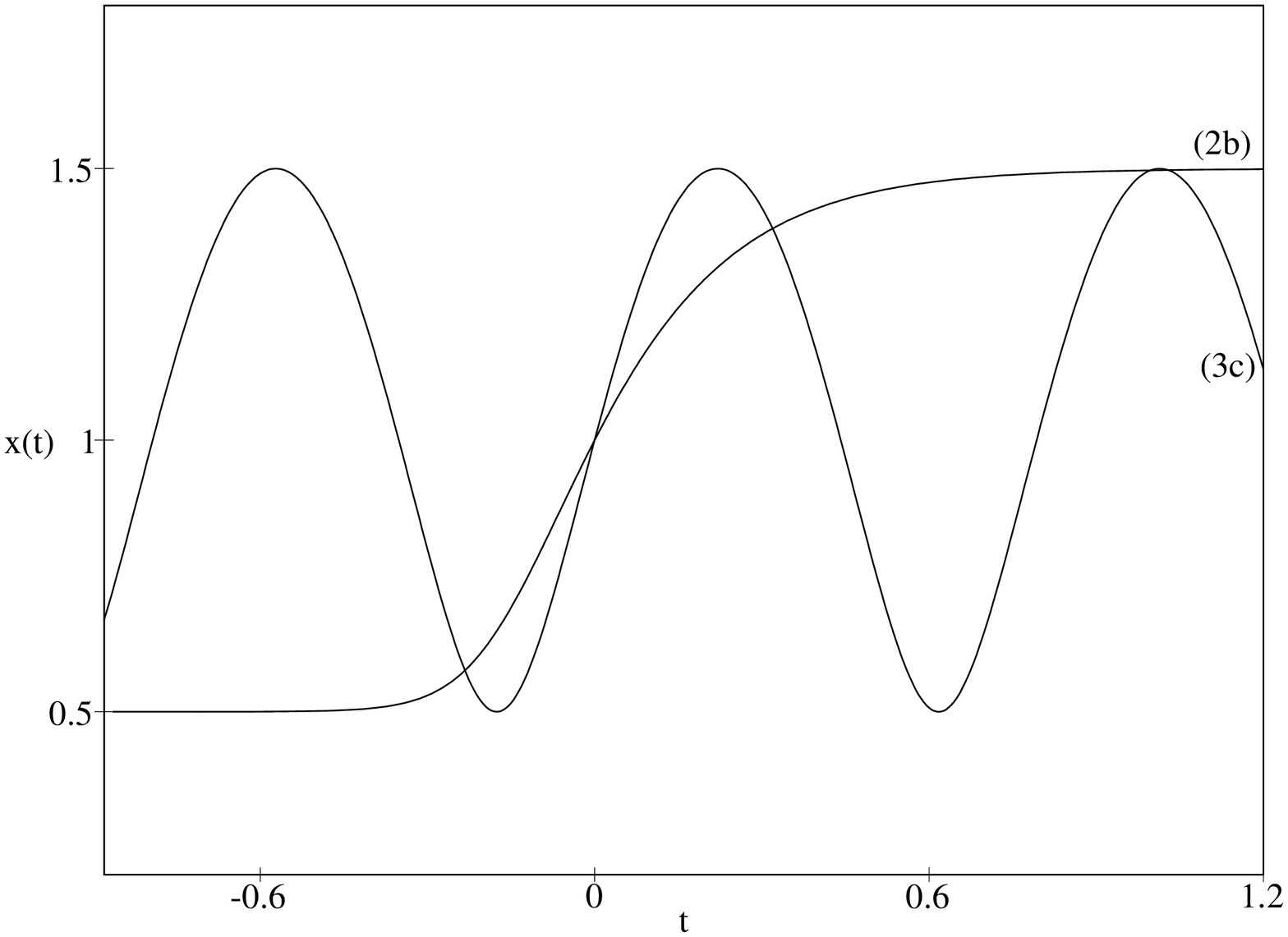}}
\resizebox{\hsize}{!}{\includegraphics[angle=-90]{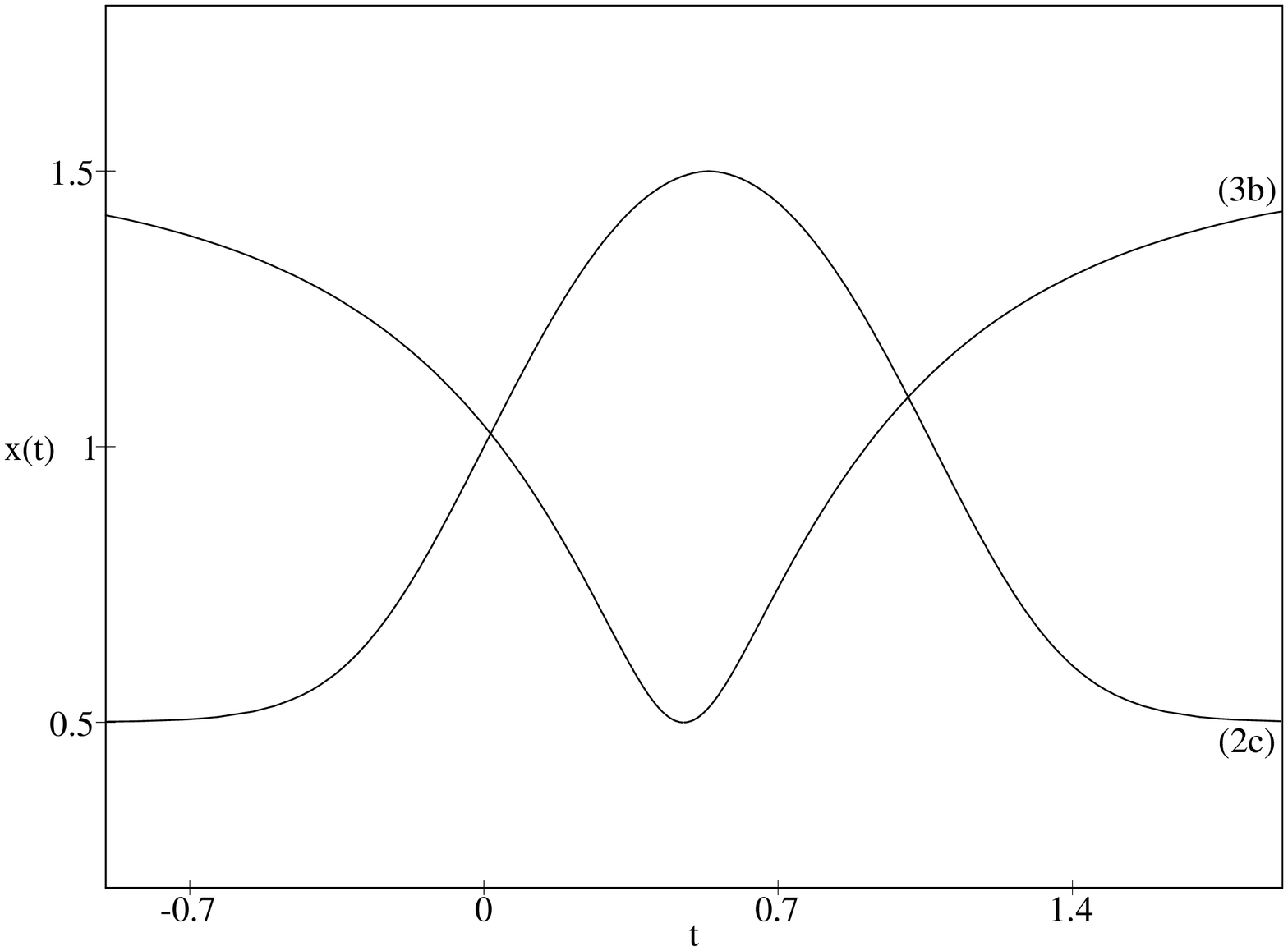}}
\caption{When fluids with negative energy are present $x(t)$ can take one of the above forms. 
Oscillating solutions are possible with two components, while the shifting-type (upper part, two asymptotic boundaries) 
as well as the hollow- and hill-type (lower part, one asymptotic boundary) require more than two fluids.}
\label{neu}
\end{figure}

Models trapped between two boundaries are shown in Fig.\,\ref{neu}. There are `shifting' (\ref{shift}), `hill-type' (\ref{hill}),
`hollow-type' (\ref{hollow}) and `oscillating' (\ref{oscill}) solutions. Apart from the periodic type
they require more than two fluid components and were not described in earlier
classifications.

The case $H_0 = 0$ leads to one more solution class: static universes with $\dot{x} = 0$ for all times. One famous
member of this class is Einstein's static universe with dust and cosmological constant.

If $H_0 <0$, then the classification remains qualitatively the same as for $H_0 > 0$.  However, solutions 
with $H_0 > 0$, that have a
symmetry-point $t_{\rm sym}$ with $x(t_{\rm sym}+t) = x(t_{\rm sym}-t)$ will be translated in time, 
when $H_0$ becomes
negative. This is the case for the original classes `big bang + max' (\ref{big bang + max}), `hill-type' (\ref{hill}), 
`bouncing' (\ref{bouncing}), `hollow-type' (\ref{hollow}) and `oscillating' (\ref{oscill}).

On the other hand, those solutions for which $\dot{x}(x) > 0$ holds for all $x \in \left[x_{\rm min},x_{\rm max}\right]$ are
reflexions of solutions with $H_0 > 0$ with respect to the $\left(t=t_0\right)$-axis. This concerns the original classes
`big bang without recollapse' (\ref{big bang}), `big bang + asymptotic upper boundary' (\ref{deSimax}), 
`asymptotic lower boundary without recollapse' (\ref{deSi}) and `shifting' (\ref{shift}).
%
%
%
%
\subsection{Repulsive and attractive fluids}
Eq.\,(\ref{exp1}) shows that a fluid is attractive, as long as $\alpha \Omega <0$, repulsive if
$\alpha \Omega >0$. If $\Omega > 0$, $\Lambda$-like fluids are repulsive, dustlike are attractive, 
whereas with $\Omega < 0$ $\Lambda$-like fluids are attractive and dustlike are repulsive. 

However, while attractive dustlike components promote a $x = 0$ singularity, 
attractive $\Lambda$-like fluids 
tend to limit the expansion to an upper boundary $x_{\rm max}$. This can be seen 
from Eq.\,(\ref{exp2}) in which each
fluid, in addition to affecting the curvature with a $H_0^2 \Omega$ term, adds a term of
the form $H_0^2 \Omega x^{\alpha} $ to the cosmic scale factor. For dustlike attractive fluids 
$\alpha < 0$ and the term is positive. Moreover, the component enforces contraction when 
the scale factor is decreasing. Conversely, for $\Lambda$-like components $\alpha > 0$ but for 
attraction $\Omega < 0$. The term is negative and its influence increases, when the universe expands. 

Repulsive dustlike fluids counteract the formation of a singularity (it depends on the effective
sum over all fluids whether a singularity will be truly avoided), 
repulsive $\Lambda$-like work against formation of an upper boundary of the scale 
factor in an expanding universe.

For a dustlike component $\Omega x^{\alpha}$ vanishes when $x \to \infty$, and the component only contributes 
to curvature. The same holds for $\Lambda$-like fluids when $x \to 0$. Therefore worldmodels, that consist of fluids
that are {\em all dustlike} or {\em all $\Lambda$-like} are dominated by curvature effects. 

Models with exclusively
dustlike, {\em attractive} fluids have necessarily a big bang. If the geometry of the model is 
\begin{itemize}
\item open, then the universe expands forever (\ref{big bang})
\item flat, then the universe expands forever and becomes asymptotically static with $x \to \infty$ (\ref{deSimax})
\item closed, then the universe passes a maximum and collapses to $x = 0$ (\ref{big bang + max}).
\end{itemize} 
Worldmodels that contain only dustlike, {\em repulsive} fluids are always open and of the bouncing type (\ref{bouncing}).

Analogously models with exclusively $\Lambda$-like, {\em repulsive} fluids expand forever, where
\begin{itemize}
\item open models contain a big bang (\ref{big bang}),
\item flat models contain an asymptotic big bang (\ref{deSi}) and
\item closed models have a minimum of the class (\ref{bouncing}).
\end{itemize}
Universes consisting of only $\Lambda$-like, {\em attractive} fluids are necessarily open and from the type (\ref{big bang + max}).
\section{Classification of two-component models}
\label{twocomp}
For classifying two-component models in the allowed range of fluid parameters, 
one needs to find the zeros of $f_2(x)$. For models with two cosmic fluids 
($\Omega_1$, $\alpha_1$) and ($\Omega_2$, $\alpha_2$)
the solution types occupy well-defined areas in the ($\Omega_1$, $\Omega_2$)-plane
for prechosen $\alpha_1$ and $\alpha_2$.  
Roots $x_{\rm c}$ of the equations $f_1(x_{\rm c}) = f_2(x_{\rm c}) = 0$ lead to the
boundary curves that separate different types of solutions. 

In the following, the parameters of state $\alpha_1$ and $\alpha_2$ as defined in 
Sect.\,\ref{multfluid} are regarded 
as arbitrary constants. Without loss of generality we assume
$\left| \alpha_1 \right| > \left| \alpha_2 \right|$.

\subsection{Transition curves}
Eqs.\,(\ref{exp1}), (\ref{exp2}) reduce to
\begin{equation}
\label{gl1}
\dot{x}^2 = H_0^2 \left( \Omega_1 x^{\alpha_1} + \Omega_2 x^{\alpha_2} + 1 - \Omega_1 - \Omega_2 \right)
\end{equation}
\begin{equation}
\label{gl2}
\ddot{x} = \frac{H_0^2}{2x} \left( \alpha_1 \Omega_1 x^{\alpha_1} + \alpha_2 \Omega_2 x^{\alpha_2} \right).
\end{equation}
Solving these equations for $f_1(x_{\rm c}) = f_2(x_{\rm c}) =0$ yields
\begin{equation}
\label{om1}
\Omega_1 = \frac{\alpha_2}{\alpha_2(1-x^{\alpha_1}_c)-\alpha_1(x^{\alpha_1-\alpha_2}_c-x^{\alpha_1}_c)}
\end{equation}
\begin{equation}
\label{om2}
\Omega_2 = \frac{- \alpha_1}{\alpha_1(x^{\alpha_2}_c-1)+\alpha_2(x^{\alpha_2-\alpha_1}_c-x^{\alpha_2}_c)}.
\end{equation}

Eqs.\,(\ref{om1}), (\ref{om2}) define a non-continuous curve in the $(\Omega_1 , \Omega_2)$
plane parameterized by the critical scale factors $x_c$. The discontinuity at $x_c =1$
divides the curve into two parts tracing models with an asymptotic upper boundary ($x_c > 1$) or an
asymptotic lower boundary ($x_c < 1$). Each branch of the curve marks a transition between 
regions with different solution types. To show this we study the case of two dustlike fluids in detail. Other
cases can be treated in an analogous way.
\subsection{Two dustlike components}
\label{twocomparg}
Recall that a dustlike fluid is attractive, as long as $\Omega > 0$. The solution 
given by a chosen pair ($\Omega_1$, $\Omega_2$) depends strongly on which 
component is attractive, which is repulsive. 

$\Omega_1>0, \Omega_2>0$:
$f_2$ has no extremum and the
type of solution depends only on the curvature. We find
\begin{equation}
\label{curvat}
\lim_{x \to \infty} f_2 = H_0^2 \left( 1 - \Omega_1 - \Omega_2 \right)
\end{equation}
and conclude that closed models ($k>0$, cf. Eq.\,(\ref{curv})) necessarily lead to a zero of $f_2(x)$ greater 
than unity. Thus they have
a maximum and recollapse into a big crunch. Open and flat ones expand forever, the latter stopping their 
expansion asymptotically. All models start with a big bang singularity.

$\Omega_1<0, \Omega_2<0$:
All models are open ($k<0$, cf. Eq.\,(\ref{curv})) and have no maximum because of
Eq.\,(\ref{curvat}) . After
collapsing from infinity and passing over a minimum they expand back to infinity, because
\begin{equation}
\label{bounce}
\lim_{x \to 0} f_2 = -\infty \ {\rm if} \ \Omega_1<0
\end{equation}
shows that there is a zero of $f_2(x)$ in the interval $\left[ 0,1 \right]$.

$\Omega_1<0, \Omega_2>0$:
From Eq.\,(\ref{bounce}) we see that such models cannot
collapse to $x = 0$. $f_1$ has only a single zero and therefore $f_2$ one extremum. According to
Eqs.\,(\ref{curvat}), (\ref{bounce}) it must be a maximum, what geometrically means that $f_2$ has no, one or two zeros,
depending on whether $f_2$ is negative, zero or positive at its maximum. 
There is at least one zero, because $f_2(t_0) > 0$ excludes a negative maximum. 
Summarizing, if the universe is open or flat it is of the bouncing 
type (\ref{bouncing}) (Eq.\,(\ref{curvat}), only one zero), but in closed worlds the solution is {\em periodic}.

$\Omega_1>0, \Omega_2<0$:
Again $f_1(x)$ has one zero, but now leading to a minimum of $f_2(x)$, because of
\begin{equation}
\label{max}
\lim_{x \to 0} f_2 = \infty \ {\rm if} \ \Omega_1>0
\end{equation}
and Eq.\,(\ref{curvat}). Therefore $f_2(x)$ has one zero greater unity if $\Omega_1 + \Omega_2 \le 1$ and 
the solutions begin with a big bang, pass through a maximum and recollapse. If $\Omega_1 + \Omega_2 > 1$
then $f_2(x)$ has no, one or two zeros, and now all possibilities are allowed in the sense of Eq.\,(\ref{conditions}). 
If there is no zero, then the universe starts with a big bang and expands forever. Two zeros lead to solutions
with one extremum, because either both are less or both 
are greater than unity. Finally, if $f_2(x)$ has only one zero, then this must be the zero of $f_1$ and, hence,
$\Omega_1$, $\Omega_2$ lie on the curve defined by Eqs.\,(\ref{om1}), (\ref{om2}). 
Thus, crossing this curve the solutions will skip from one type to the other.

\begin{figure}
\resizebox{\hsize}{!}{\includegraphics[angle=-90]{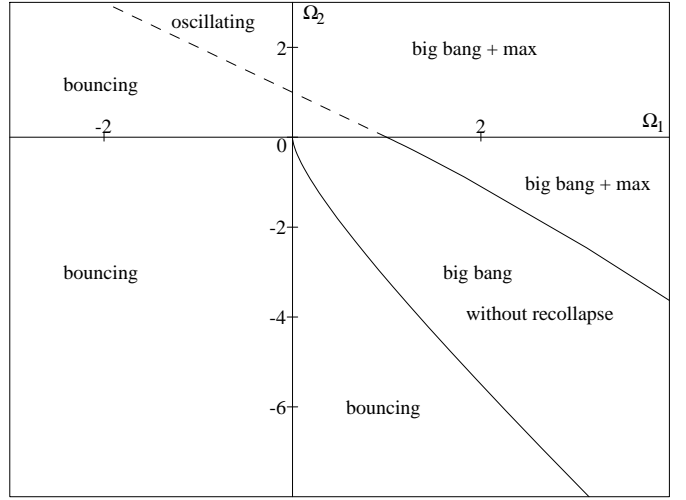}}
\caption{The density-parameter plane for two dustlike fluids with ($\Omega_1$, $\alpha_1=-1.7$), ($\Omega_2$, $\alpha_2=-1$).
The type of solution depends on the strength of each component. Models on the dashed line are flat.}
\label{class1}
\end{figure}

Fig.\,\ref{class1} summarizes the results for specific values $\alpha_1=-1.7$ $\alpha_2=-1$.
If $\alpha_1$ becomes smaller, or $\alpha_2$ becomes larger, then the transition curves
become steeper. In this sense, one may refer to $\alpha$ as a parameter characterizing the 
strength of the attraction or repulsion of a fluid. The qualitative properties of the transition curves 
do not change under variation of $\alpha_1$ and $\alpha_2$.

\subsection{The general case}
The general case of two arbitrary components follows analogously.

\begin{figure}
\resizebox{\hsize}{!}{\includegraphics[angle=-90]{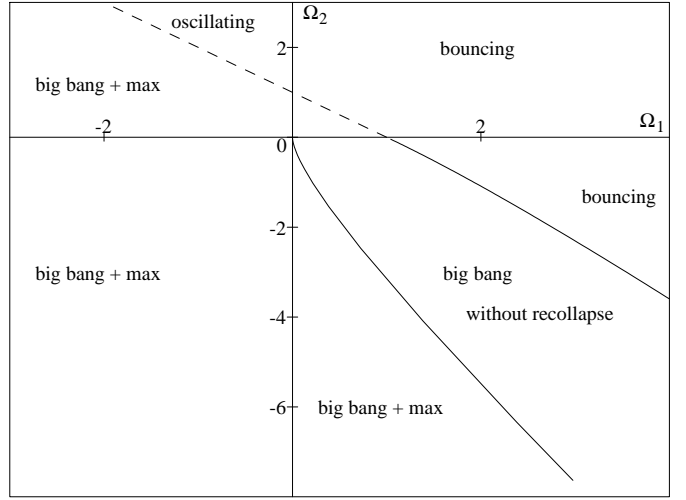}}
\caption{The density-parameter plane for two $\Lambda$-like fluids, ($\Omega_1$, $\alpha_1=1.7$), ($\Omega_2$, $\alpha_2=1$). The roles of positive and 
negative energy densities compared to Fig.\,\ref{class1} have interchanged. As in Fig.\,\ref{class1} flat models are 
related to the dashed line.} 
\label{class2}
\end{figure}

Fig.\,\ref{class2} shows the regions of solution types for two $\Lambda$-like fluids. 
The assumption $\alpha_1 > \alpha_2$
enforces the transition curve to occur in the lower right quadrant. The curves look
similar to those of Fig.\,\ref{class1}, however the roles of positive and negative energy 
densities have interchanged so that the domains `bouncing' and `big bang+max' flipped
over. This
illustrates the opposite behavior of dustlike and $\Lambda$-like fluids.
The transition curves will get steeper, when either $\alpha_1$ becomes larger,
or $\alpha_2$ becomes smaller (or both together).

\begin{figure}
\resizebox{\hsize}{!}{\includegraphics[angle=-90]{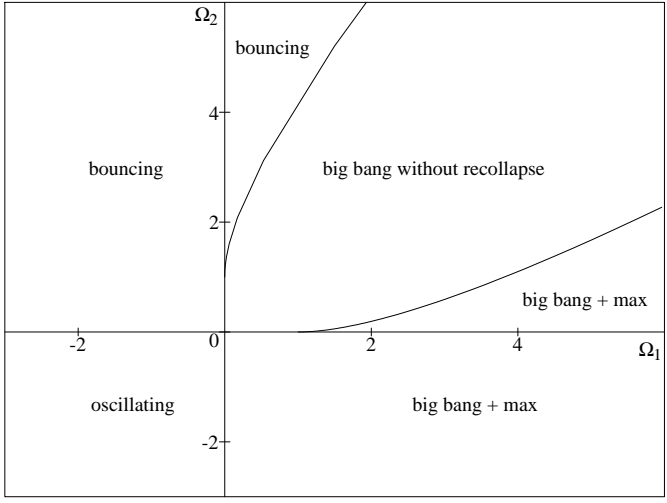}}
\caption{The case of one dustlike fluid ($\Omega_1$, $\alpha_1=-1.4$) and one $\Lambda$-like
fluid ($\Omega_2$, $\alpha_2=1.5$) shown here is similar to that of universes containing dust and a 
$\Lambda$-term. The transition curves become steeper if $\alpha_1$ or $\alpha_2$ decreases.}
\label{class3}
\end{figure}

In Fig.\,\ref{class3} a dustlike fluid ($x$-axis) is mixed with a $\Lambda$-like. Now the transition 
curves have moved into
the upper right quadrant. This reflects the fact that fluids with different signs in the parameter of state act contrary
when the signs of the densities are equal in contrast to the previous cases.
%
%
%
\section{Models with one exotic component}
\label{more}
Recent observations of supernovae of type Ia (Perlmutter et al. 1998) suggest the presence of at least one
repulsive fluid in addition to the cosmic radiation background (CMBR)
$\left( \Omega_{\rm \gamma} \ , \ \alpha_{\gamma}=-2 \right)$ and pressureless (incoherent) matter (dust)
$\left( \Omega_{\rm M}, \alpha_{\rm M}=-1 \right)$. For $\Omega_{\rm \gamma}$ we can assume 
$\Omega_{\rm \gamma} \in \left[0,1\right]$. Incoherent matter encompasses `ordinary' baryonic matter 
and possibly what is called cold dark matter (see Sect.\,\ref{disc} for details). Due to our ignorance of the
physics and amount of dark matter it seems appropriate for classification not to restrict the value 
of $\Omega_{\rm M}$. 
The incoherent component is attractive, as long as its density is positive. 

For three components Eqs.\,(\ref{exp1}), (\ref{exp2}) read
\begin{equation}
\label{exp13}
\dot{x}^2 = H_0^2 \left( \frac{\Omega_{\rm M}}{x} + \frac {\Omega_{\rm \gamma}}{x^2} + \Omega x^{\alpha}
  + 1 -  \Omega_{\rm M} - \Omega_{\rm \gamma} - \Omega \right)
\end{equation}
\begin{equation}
\label{exp23}
\ddot{x} = \frac {H_0^2}{2x} \left(-\frac{\Omega_{\rm M}}{x} - \frac {2 \Omega_{\rm \gamma}}{x^2} 
   + \alpha \Omega x^{\alpha} \right)
\end{equation}
As in Sect.\,\ref{twocomp} we need to find out simultaneous zeros of $f_2$ and $f_1$ for 
classifying 
the solutions. However, we now regard the parameters of the `third' component,
parameter of state $\alpha$ and normalized density $\Omega$
as independent variables. The corresponding equations for $(\Omega,\ \alpha)$ 
cannot be solved analytically for $\alpha$. 

By transforming Eqs.\,(\ref{exp13}), (\ref{exp23}) into
\begin{equation}
\label{exp33}
\Omega(\alpha,x_c) = \frac {\Omega_{\rm M}/x_c + 2\Omega_{\rm \gamma}/x^2_c}{\alpha x^{\alpha}_c}
\end{equation}
\begin{eqnarray}
\label{exp34}
\frac {x^{\alpha}_c - 1}{\alpha x^{\alpha+2}_c} \, (\Omega_{\rm M}x_c + 2\Omega_{\rm \gamma}) &+& \nonumber \\
 \frac {\Omega_{\rm M}}{x_c} + \frac {\Omega_{\rm \gamma}}{x^2_c} 
&+& 1 - \Omega_{\rm M} - \Omega_{\rm \gamma} = 0
\end{eqnarray}
one can proceed to solve Eq.\,(\ref{exp34}) numerically for $\alpha(x_c)$. Inserting this result 
into Eq.\,(\ref{exp33}) yields $\Omega(\alpha(x_c),x_c)$. 

\begin{figure}
\resizebox{\hsize}{!}{\includegraphics[angle=-90]{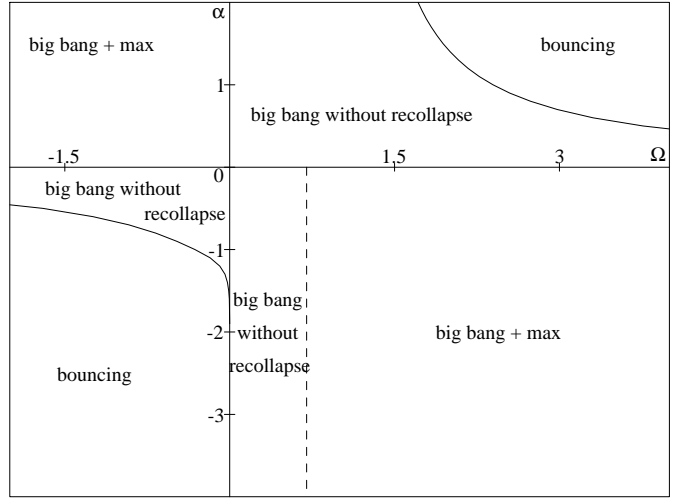}}
\caption{Possible solution types for three components, among which 
dust ($\Omega_{\rm M} = 0.3$) and radiation ($\Omega_{\rm \gamma} = 5.9\,10^{-5}$) are fixed. $\Omega$ and
$\alpha$ are the parameters of a third component. Models on the dashed line are flat.}
\label{class4}
\end{figure}

If $\Omega_{\rm M} > 0$ and $\Omega_{\rm M} + \Omega_{\rm \gamma} < 1$, then
Eq.\,(\ref{exp34}) has no solution for
$x_c>1$ and the curve $\Omega(x_c)=\Omega(\alpha(x_c),x_c)$ describes models with an asymptotic
lower boundary, separating bouncing solutions from those with big bang, that have no maximum. 
It splits into two branches for $\alpha >0$ and $\alpha<0$, respectively.
Fig.\,\ref{class4} shows the distribution of solution classes in the $(\Omega$, $\alpha)$ plane for this case.
The individual domains arise from arguments comparable to that in Sect.\,\ref{twocomparg}. Due
to the restrictions we made for $\Omega_{\rm M}$ and $\Omega_{\rm \gamma}$ none of the new solution
types of Fig.\,\ref{neu} arises here. 

\begin{figure}
\resizebox{\hsize}{!}{\includegraphics[angle=-90]{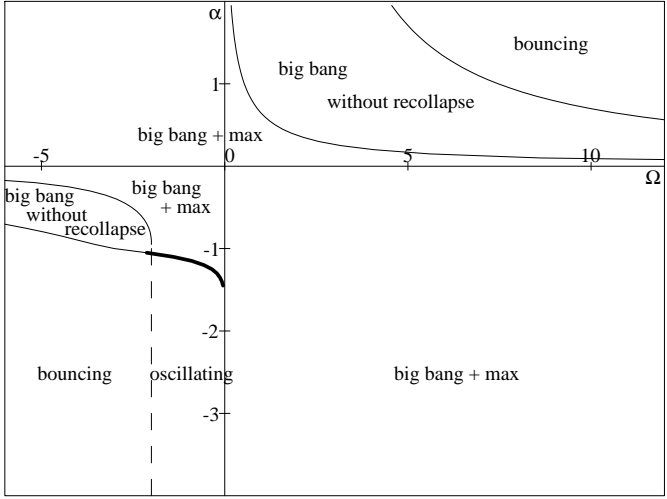}}
\caption{Classification of solution types for three components, $\Omega_{\rm \gamma} = 5.9\,10^{-5}$, 
$\Omega_{\rm M} > 1 - \Omega_{\rm \gamma}$ and an arbitrary third component ($\Omega$, $\alpha$). 
The models on the dashed line are flat, that on the bold part of the transition curve have hill-type solutions.}
\label{class4a}
\end{figure}

In Fig.\,\ref{class4a} the result of classification is shown for $\Omega_{\rm M} > 1 - \Omega_{\rm \gamma}$. The
greater value of $\Omega_{\rm M}$ increases the number of posssible solution types. By the differences of
Figs.\,\ref{class4} and \ref{class4a} the general fact is expressed, that a set of dustlike fluids $\Omega_i > 0$ obeying
$\sum_i \Omega_i < 1$ cannot lead to a maximum of $x(t)$ without further attractive components. Therefore Fig.\,\ref{class4a}
shows solutions with a maximum, where the third component is repulsive (upper right quadrant), that are caused
by the greater value of $\Omega_{\rm M}$. Accordingly, there are two different types of transition curves (as in
Figs.\,\ref{class1}-\ref{class3}). Between `bouncing' and `big bang without recollapse' the solutions have an 
asymptotic lower boundary, between `big bang without recollapse' and `big bang + max' they have an
asymptotic upper boundary.

\begin{figure}
\resizebox{\hsize}{!}{\includegraphics[angle=-90]{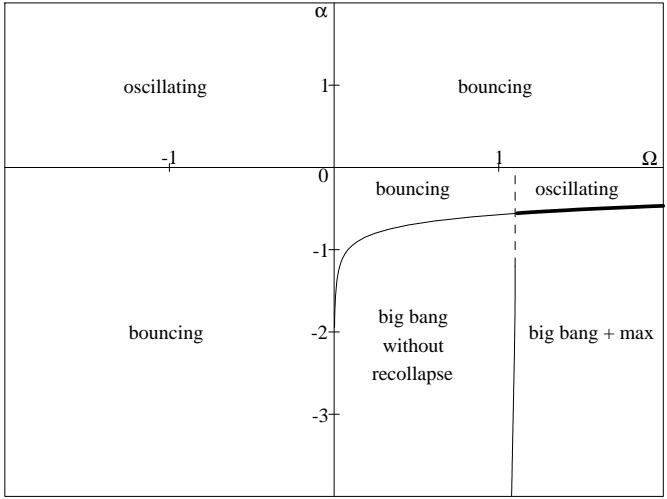}}
\caption{Solutions with three components, $\Omega_{\rm \gamma} = 5.9\,10^{-5}$, $\Omega_{\rm M} < -0.015$ and no
restricitions for the third component ($\Omega$, $\alpha$). flat models are on the dashed line, hollow-type solutions
on the bold part of the transition curve.}
\label{class4b}
\end{figure}

Fig.\,\ref{class4b} shows the solution types, when $\Omega_{\rm M} < \Omega_{\rm crit}$, where 
$\Omega_{\rm crit}$ represents the critical dust density that excludes a big bang in dust-radiation-models. 
$\Omega_{\rm crit}$ depends on $\Omega_{\rm \gamma}$, for example with
$\Omega_{\rm \gamma} = 5.9\,10^{-5}$ (see Sect.\,\ref{obs}) one finds $\Omega_{\rm crit} = -0.015$. 

For the transition between different
solution types the same holds as for Fig.\,\ref{class4a}.
%
%
%
\section{Models with two exotic components}
An increasing number of components makes the classification more and more difficult. When the geometry of
$f_2(x)$ becomes richer in variants, e.g. if there are more extrema, then the method of classification
introduced above is no longer useful, the simultaneous zeros of $f_2(x)$ and $f_1(x)$ are in general no longer
distinctive for different solution types. However, one can get the solution type directly from $f_2(x)$ in any
case (see Sect.\,\ref{dynamics}).

\begin{figure}
\resizebox{\hsize}{!}{\includegraphics[angle=-90]{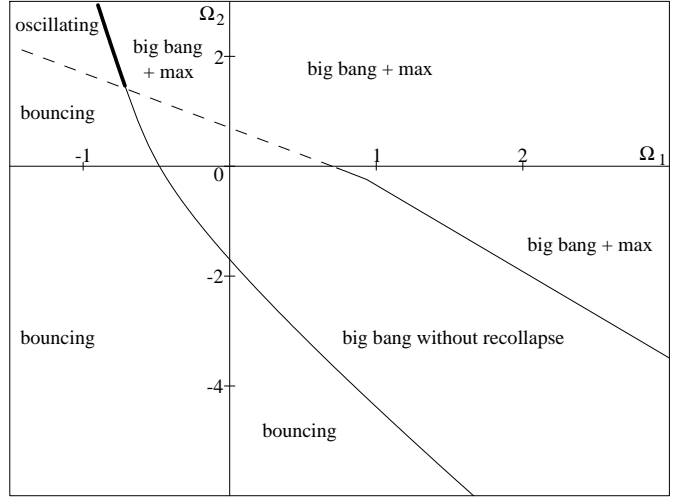}}
\caption{The possible solutions in four component models with dust, radiation and 
two dustlike, exotic components ($\Omega_1$, $\alpha_1 = -0.9$), ($\Omega_2$, $\alpha_1 = -0.5$). 
Solutions on the transition curves have an asymptotic lower boundary, on the bold part they have also a maximum 
(hill-type). Models on the dashed line are flat.}
\label{class5}
\end{figure}

\begin{figure}
\resizebox{\hsize}{!}{\includegraphics[angle=-90]{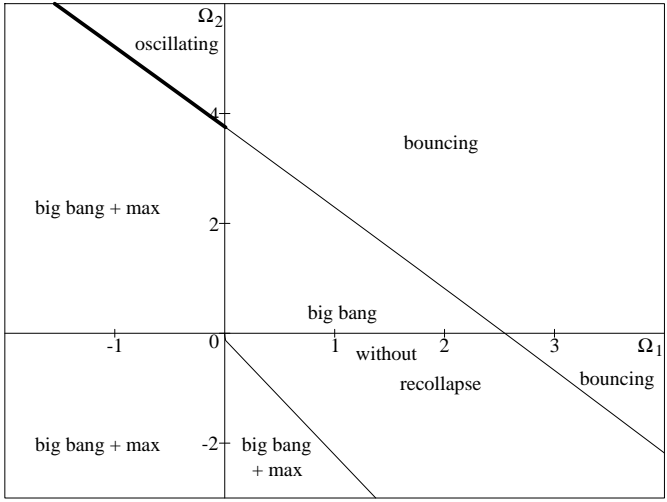}}
\caption{Worldmodels with dust, radiation and two $\Lambda$-like, exotic fluids 
($\Omega_1$, $\alpha_1 = 0.9$), ($\Omega_2$, $\alpha_1 = 0.5$). In contrast to
Fig.\,\ref{class5} and corresponding to Fig.\,\ref{class2} the effect of positive and negative energy
densities have interchanged. For the solutions on the curves the same as for Fig.\,\ref{class5} holds.}
\label{class6}
\end{figure}

\begin{figure}
\resizebox{\hsize}{!}{\includegraphics[angle=-90]{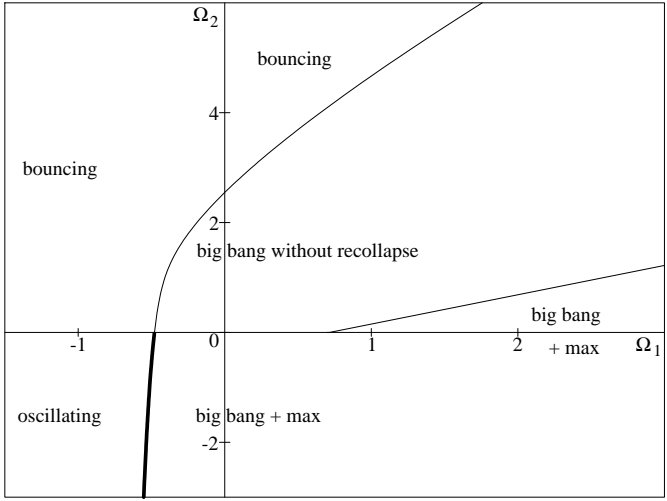}}
\caption{The domains of different solution types for one 
dustlike ($\Omega_1$, $\alpha_1 =-0.9$) and one $\Lambda$-like fluid ($\Omega_2$, $\alpha_2 =0.9$) as exotic
additions to dust and radiation. 
The oscillating and recollapsing big-bounce models
are separated by hill-type solutions (bold part of the transition curve).}
\label{class7}
\end{figure}

As in Sect.\,\ref{twocomp} we represent the solution types in the $(\Omega_1$, $\Omega_2)$-plane of the densities of
the additional components $(\Omega_1$, $\alpha_1)$, $(\Omega_2$, $\alpha_2)$.
They now look like in Figs.\,\ref{class5}-\ref{class7}, for which we have adopted 
$\Omega_{\rm M} = 0.3$ and $\Omega_{\rm \gamma} = 5.9\,10^{-5}$ (see Sect.\,\ref{obs}). However, there is no
qualitative change in the classification as long as $\Omega_{\rm M} + \Omega_{\rm \gamma} < 1$ and
both densities are positive. Each of the figures corresponds
 to one of Figs.\,\ref{class1}-\ref{class3}
according to the combination of dustlike and $\Lambda$-like fluids as additional components.
As a difference, the transition curves are no longer restricted to one quadrant (though this can happen with suitable
values of $\alpha_1$ and $\alpha_2$). The bold parts of the curves mark the border
between oscillating solutions and those with big bang and maximum. They represent `hill-type' solutions.
The transition curves between `bouncing' and `big bang without recollapse' belong to solutions with an 
asymptotic lower boundary, those between `big bang without recollapse' and `big bang + max' belong to solutions
with an asymptotic upper boundary.

With increasing $\Omega_{\rm M}$ or $\Omega_{\rm \gamma}$ the regions of big-bang 
models in Figs.\,\ref{class5}-\ref{class7}
(that do not occur in the corresponding Figs.\,\ref{class1}-\ref{class3}) grow because of the greater amount of
attractive energy. The transition curves of Fig.\,\ref{class5}
and \ref{class7} move to the left, those of Fig.\,\ref{class6} to the right.

If one of the component densities vanishes (let us formally set $\Omega_{\rm M} := 0$), then Figs.\,\ref{class5} and 
\ref{class7} still hold. We can formally reidentify $\Omega_1 = \Omega_{\rm M}$ ($\alpha_1 = -1$) and then 
the figures represent the case of three component models with dust, radiation and {\em one} exotic fluid 
(which is described by $\Omega_2$ and $\alpha_2$). As a difference to Figs.\,\ref{class4} and \ref{class4a}, now 
the densities $\Omega_{\rm M}$ and $\Omega$ are the independent parameters.
%
%
%
%
%
\section{Supernova constraints}
\label{obs}
Employing standard least-square techniques we derived best-fit parameters for a number of
models compared to the supernova data set of Perlmutter et al. (1999) (for details
see Thomas 2000). We found that with
a dust component $\Omega_{\rm M} > 0$ representing `ordinary' and cold dark matter in the universe the observations
force a repulsive, $\Lambda$-like component ($\Omega$, $\alpha$) in our framework of description. Our best fit yields
$$
\Omega_{\rm M} = 0.86 \pm 0.58 \ , \ \Omega = 1.38 \pm 0.73 \ , \ \alpha = 2 \pm 1.9.
$$
According to Padmanabhan (1993) we write
$$
\Omega_{\rm \gamma} = 2.5 \,10^{-5}/h^2, \ \ {\rm with} \ H_0 = h \, 100 \frac{{\rm km/s}}{{\rm Mpc}}
$$ for the CMBR component. Using $h = 0.65$ one obtains $\Omega_{\rm \gamma} = 5.9 \,10^{-5}$. With
this value the influence of the CMBR component is negligible.

The great uncertainty in the value of the parameter of state $\alpha$ expresses that nearly any $\Lambda$-like,
repulsive fluid can be fit to the data. When $\Omega_{\rm M}$ decreases then $\Omega$ grows, while $\alpha$
decreases, too. For very low values of $\Omega_{\rm M}$, e.g. about $\Omega_{\rm M} \approx 0.03$, close to the 
baryon density  predicted
by big-bang nucleosynthesis, one obtains $\alpha \approx 0.5$.

Among the models with a $\Lambda$-like repulsive fluid that fit to the data there are solutions without a big-bang
beginning. These solutions lead to $\alpha \approx 0.5$. A necessary 
condition for nucleosynthesis comparable to that of big-bang nucleosynthesis is to reach $T \sim 10^9$ K. 
Therefore one may restrict  such world models to obey
$x_{\rm min} < 10^{-9}$. To fulfill this condition $\Omega_{\rm M}$ and $\Omega$ have to be appropriately fine tuned
which can be easily done. It is beyond the scope of the present work to check detailed
nucleosynthesis scenarios for such exotic models, which would also depend on baryonic density and time scales
at the nucleosynthesis epoch.

\begin{figure}
\resizebox{\hsize}{!}{\includegraphics[angle=-90]{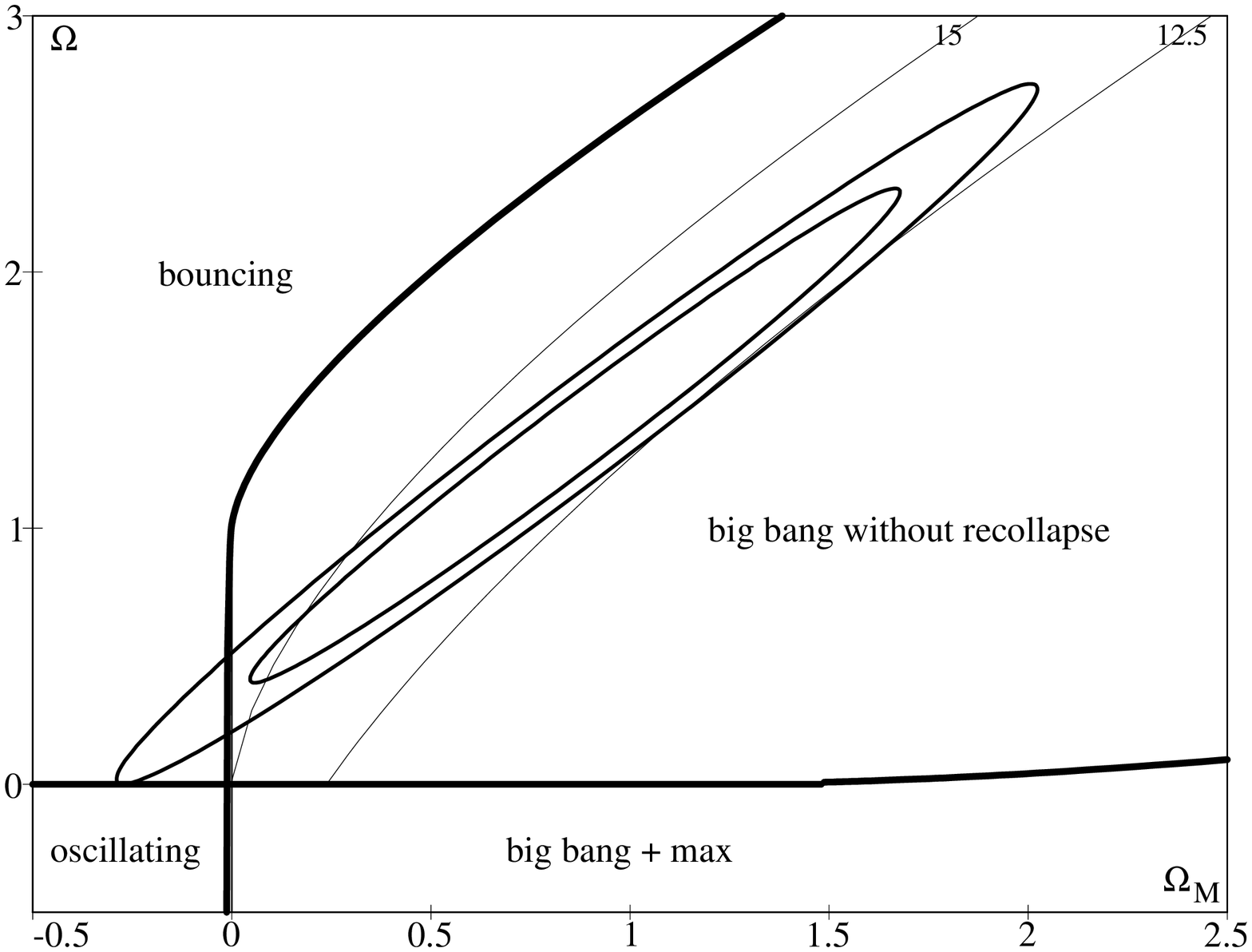}}
\resizebox{\hsize}{!}{\includegraphics[angle=-90]{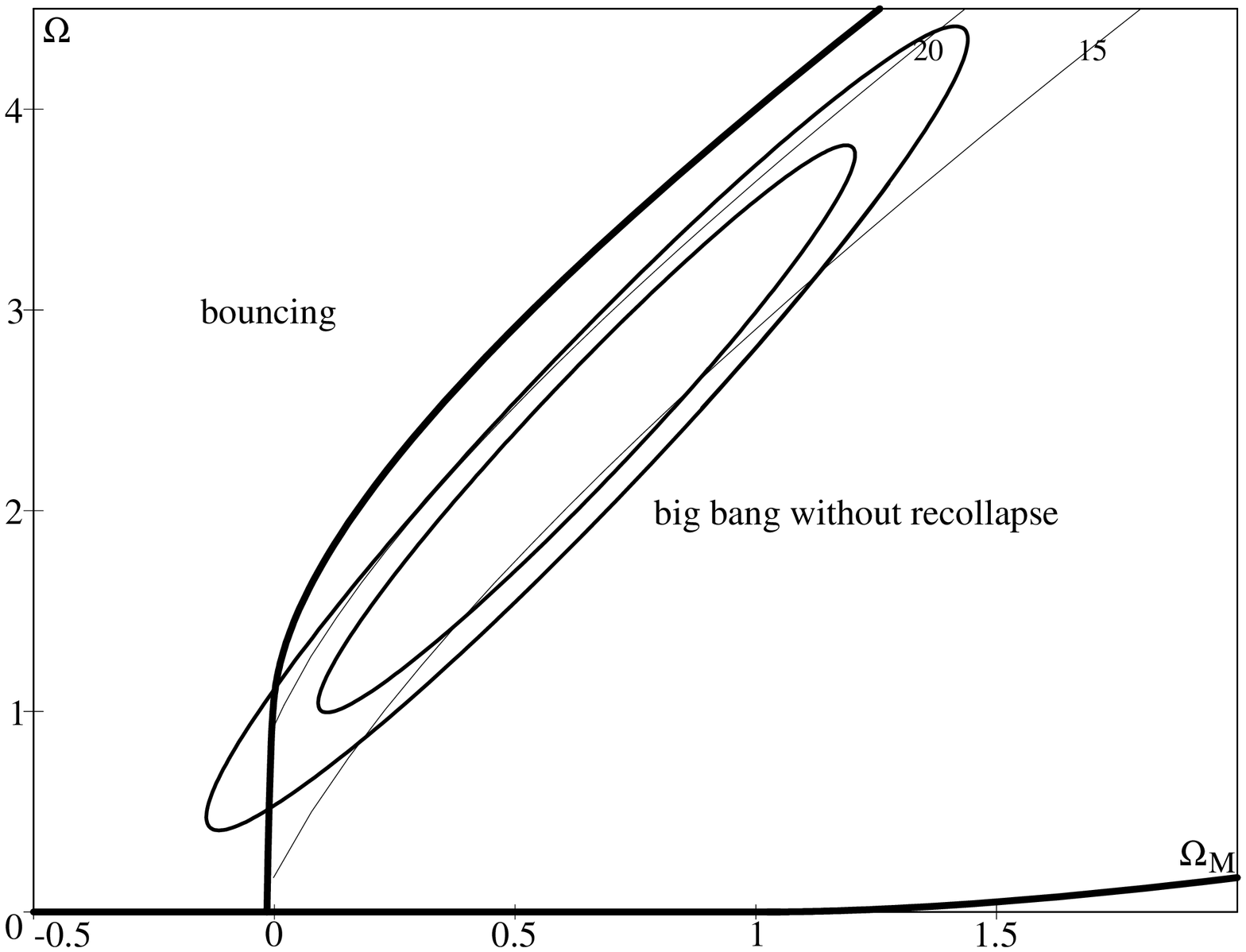}}
\caption{This figure shows 68.3\% and 90\% confidence regions for world models that have been fit to the SN Ia
data set from Perlmutter et al. (1999). The upper panel corresponds to 2-component models
consisting of dust ($\Omega_{\rm M}$) and
a $\Lambda$-fluid (density $\Omega$ and state parameter $\alpha=2$), whereas a composition of dust plus a 
domain-wall like fluid (density $\Omega$ and state parameter $\alpha=1$) has been utilized in the lower panel. The 
fat lines separate models of different solution type, the thin lines represent models with constant age (in Gyr).}
\label{fitpics}
\end{figure}

A world model can only be considered as a successful fit if its age $t_0$ exceeds 
the ages of the oldest known objects. 
In Fig. \,\ref{fitpics} we plotted thin lines of constant age in an 
$\Omega_M$-$\Omega$-diagram for two-component models.
The state parameters $\alpha$ of the exotic component with density $\Omega$ are 
$\alpha=2$ ($\Lambda$-fluid) in
the upper panel of the figure and $\alpha=1$ (domain-wall like fluid) in the lower panel. The 90\%
confidence regions of the fits given in the diagrams are roughly encompassed by 
iso-age lines (for $h = 0.65$)
corresponding to 12.5 and 15 Gyrs (for $\alpha=2$) or 15 and 20 Gyrs (for 
$\alpha=1$), reveiling
the trend to larger $t_0$ for decreasing $\alpha$. When approaching the fat 
vertical-diagonal line from below, 
the world ages would increase towards infinity at this line, which separates 
big-bang models from
bouncing models.

Stars in old globular clusters are considered to be among the oldest objects 
whose age
can be determined in a straightforward manner provided that calibration 
uncertainties (most importantly those
of the RR Lyrae absolute magnitudes)
are narrowed down. A recent revision of the ages of globular clusters yielded
$11.5 \pm 1.3$ Gyrs (Chaboyer et al. 1998). Allowing a formation time of globular 
cluster
stars of at least $\sim 1$ Gyr leads to the condition $t_0 > 12.5$ Gyrs, which is 
fulfilled by the
models inside the 90\% error ellipse in Fig.~\ref{fitpics}.

When taking into account recent results from the CMBR first acoustic peak which require a 
nearly flat universe (de Bernardis et al. 2000) we
find a tendency $\alpha \to 2$. Consequently this favours the repulsive component to be the $\Lambda$-fluid. Our
best fit with a flat model ($k=0$) yields
$$
\Omega_{\rm M} = 0.30 \pm 0.04 \ , \ \Omega = 1 - \Omega_{\rm M} \ , \ \alpha = 2 \pm 0.04.
$$

$\Omega_{\rm M} = 0.3$ is in agreement with constraints from peculiar motion (Zehavi \& Dekel 1999).
Additional exotic fluids do not improve the goodness of fit and may therefore be considered as unnecessary. However,
due to the increasing number of free parameters with each additional fluid one can fit models to the data
that have for instance no big bang. Possible fits include models with $x_{\rm min} < 10^{-9}$, $\Omega_{\rm M} = 0.3$, 
$\Omega_{\rm \gamma} = 5.9 \,10^{-5}$ that are flat and fit to the data with two exotic repulsive fluids, one of them
$\Lambda$-like, the other dustlike.
%
%
%
\section{Discussion and conclusions}
\label{disc}
We have extended available classification schemes for CP world models by allowing for 
generalized ideal fluids with ${\cal P} = \left( \gamma - 1 \right) \epsilon$, $\gamma \in \left[ 0,2 \right]$.
In particular, we have not restricted ourselves to only encompass incoherent matter (`dust')
and a $\Lambda$-field, though
these are included as special cases. Other fields beyond $\Lambda$, which are repulsive due to
sufficient negative pressure, are commonly applied
in theories of the early universe (Kolb \& Turner 1990). Since supernova Ia data
require a currently acting repulsive component  (e.g. Perlmutter et al. 1999), whose physical nature
is unknown, a more general consideration of possible solutions is in order. It has become popular to 
employ the quintessence {\em ansatz}, which implies a varying equation of state as 
compared to our approach (e.g. Caldwell et al. 1998; Wang et al. 1999). This approach is less suitable
for classification purposes.

We here utilized a multifluid approach with fixed equations of state (as defended by
Turner \& White 1997), but generously allowed for negative energy densities in addition to the
possibility of negative pressure.  Such negative energies need not necessarily be due to matter
fields. They could, for instance, arise in the context of 
turbulence (cf. Marochnik et al. 1975a, b).  From the beginning of relativistic cosmology
negative energy densities were implicitely considered by allowing for the possibility of a negative 
cosmological constant $\Lambda$ (see Sect.\,2.1). 

General relativity for itself contains no strong constraints on the contents of its energy-moment tensor.
Hawking \& Ellis (1974) summarized how singularity theorems can be derived if one imposes certain 
`reasonable' {\em energy conditions}. These energy conditions are not anchored in deeper theory, but are
rather assumptions that are apparently fulfilled by `normal' macroscopic matter. Quantized matter fields
can violate the energy conditions (e.g. Parker \& Fulling 1973). The occasionally occuring 
`age problem' of CP world models, imposed by an uncomfortably large $H_0$ combined with a large age of the
oldest known stars (often from globular cluster fits), leads easily to a necessary violation of the
energy conditions if one sticks to CP universes (Visser 1997). Part of the exotic components considered in our 
classification can also violate the weak and dominant energy conditions; the strong energy condition
is violated by any repulsive component. 
 
As our analysis shows, negative energy densities as well as negative pressures 
do not lead to mathematical difficulties under the condition of homogenity and isotropy. 
The transition curves smoothly cross
the borders of the quadrants. Instead they make four new types of solutions possible, that were not yet discussed. 
There are the `hill-type', `hollow-type', `oscillating' and `shifting-type' solutions (Fig.\, 1), all free of a big bang and 
therefore without a beginning at finite times in the past.

In universes with up to two general-fluid components none of these new solution types appear except the
`oscillating' type, and 
the relations in the density plane do not qualitatively differ from that of universes 
with dust and $\Lambda$. 

Essentially, fits to the luminosity distances as a function of redshift to observed
type-Ia supernovae suggest the presence of at least one
{\em repulsive} cosmic component in addition to the `ordinary' matter and radiation
content (Perlmutter et al. 1999). We demonstrated that
there are models with repulsive components which do
not necessarily start with a big bang. However, if accepting the well-known
successes of big-bang models, which include the explanation of the abundances of
light elements and the CMBR, such bouncing models 
should be tuned to be successful in this regard as well.
This requires at least that the scale factor falls short of 
$x \approx 10^{-9}$ at the bouncing epoch to yield temperatures of $\sim 10^9$ K. 
This can be easily achieved with one additional exotic fluid. However, such a fit would lead to another
`fine-tuning' problem due to the lack of any known physical reason for the value
of the tuning parameter. Since there is, moreover, no other reason for invoking such
components we follow Occams razor to confine ourselves on the minimal currently
required number of cosmic fluids and do not pursue such possibilities further.
It should nevertheless be emphasized that
the so-called inevitability of big-bang models (e.g. B{\"o}rner \& Ehlers 1988) rests 
on such assumptions that the universe only consists of a minimal number 
of $\Omega>0$ fluids. 

With $\Omega_{\rm M}$ and $\Omega_{\rm \gamma}$ as above and {\em one} repulsive 
component ($\Omega$, $\alpha$) we find that the universe 
will expand forever and has started with a big bang in almost all models that fit the supernova data. At the
transition between models with big bang and those without, the age of the universe smoothly grows to
infinity.
An exotic $\Omega < 0$ component does not yield models in good agreement 
with the observations. Our fits showed that
 instead a $\Lambda$-like fluid with $\Omega > 0$ seems very likely. In particular, if one relies
additionally on the first acoustic peak in the CMBR the universe should be nearly flat. In this case the 
$\Lambda$-fluid itself fits best as repulsive component.
None of the new solutions appears in this context.
However, if assuming four components one may also  
tune bouncing flat solutions to supernova and nucleosynthesis conditions.

Since the current Supernova Ia data reaching to $z \approx 1$ are compatible with a large
number of multifluid models further observations are needed to reduce the uncertainty in determining
empirically the cosmic equation of state. The observational relation has to be improved by more
accurate or many more observations (tightening the mean relation) and extended to higher
redshifts where the model curves tend to diverge more.
\begin{acknowledgements}
The authors are indebted to Rolf Chini for support and encouragement.
\end{acknowledgements}

\end{document}